# Functional brain networks reveal the existence of cognitive reserve and the interplay between network topology and dynamics

Johann H. Martínez *, María E. López *, Pedro Ariza, Mario Chavez, José A. Pineda- Pardo, David López-Sanz, Pedro Gil, Fernando Maestú and Javier M. Buldú.

[1] INSERM, Institut du Cerveau et de la Moelle Epinière (ICM). Hôpital Pitié Salpétrière, Paris, France
[2] Laboratory of Biological Networks, Center for Biomedical Technology (CTB), Madrid, Spain
[3] Department of Basic Psychology II, Complutense University of Madrid, Madrid, Spain
[4] 7CNRS-UMR 7225, Hôpital Pitié-Salpetrière, Paris, France.
[5] Centro Integral de Neurociencias AC (CINAC), HM Puerta del Sur, Madrid, Spain
[6] CEU San Pablo University, Madrid, Spain.
[7] Laboratory of Cognitive and Computational Neuroscience (UCM-UPM), Center for Biomedical Technology (CTB), Madrid, Spain
[8] Institute of Sanitary Investigation [IdISSC], San Carlos University Hospital, Madrid, Spain
[9] Geriatrics Department, San Carlos University Hospital, Madrid, Spain.
[10] Complex Systems Group, Universidad Rey Juan Carlos, Madrid, Spain

* These authors have contributed equally and must be considered as the first authors

ABSTRACT

We investigated how the organization of functional brain networks was related to cognitive reserve (CR) during a memory task in healthy aging. We obtained the magnetoencephalographic functional networks of 20 elders with a high or low CR level to analyse the differences at network features. We reported a negative correlation between synchronization of the whole network and CR, and observed differences both at the node and at the network level in: the average shortest path and the network outreach. Individuals with high CR required functional networks with lower links to successfully carry out the memory task. These results may indicate that those individuals with low CR level exhibited a dual pattern of compensation and network impairment, since their functioning was more energetically costly to perform the task as the high CR group. Additionally, we evaluated how the dynamical properties of the different brain regions were correlated to the network parameters obtaining that entropy was positively correlated with the strength and clustering coefficient, while complexity behaved conversely. Consequently, highly connected nodes of the functional networks showed a more stochastic and less complex signal. We consider that network approach may be a relevant tool to better understand brain functioning in aging.

INTRODUCTION

In recent years, there is an increased interest in studying lifestyle factors that may help to achieve a successful aging. These factors, such as educational level, leisure activities or intelligence quotient (IQ) have been related to the concept of reserve, which emerges from the absence of a direct relationship between the severity of brain pathology and its clinical manifestations [1].

Two non-mutually exclusive models have been proposed to address these counterintuitive observations [1]: a passive or brain reserve (BR) model, which considers that the amount of neural substrate (e.g. the brain size or neuron quantity) determines the threshold beyond which clinical and functional deficits emerge [2,3]; and an active or cognitive reserve (CR) model, that refers to the ability to use the existent resources as flexible and efficiently as possible during the execution of cognitive tasks (neural reserve) or when coping with brain pathology (neural compensation) [4]. Healthy aging and dementia, especially Alzheimer's disease (AD), have been considered optimal models to understand the role of reserve [4–6].

Among lifestyle variables, the educational level (alone or in combination with occupational attainment) is one of the most studied and important CR proxies. In fact, some studies have considered the educational level as a "protective" factor by reducing the incidence of dementia [5], and its pathological consequences in AD patients [7,8]. However, recent studies did not find this effect [9,10], suggesting that CR actually may have a "masking" effect.

One of the new approaches to explore how brain functioning may be modulated by CR is functional network organization (based on functional connectivity; FC), although there is still scarcity of studies focused on this field. Solé-Padullés et al. [11] described that healthy elders with higher CR showed reduced functional magnetic resonance imaging (fMRI) activation during cognitive processing. In the same line, a previous study from our group [12] observed that healthy elders with a high CR score in comparison with those with low CR score, exhibited less magnetoencephalographic (MEG) functional connectivity during the performance of a memory task. These results may suggest that higher CR is related to more effective use of cerebral resources although this hypothesis has not been tested yet.

In addition, FC may be used to explore brain's organization in terms of structural and functional networks, by the application of different methodologies coming from Network Science [13]. This kind of analysis can introduce a different perspective to better understand how CR modules functional resilience in healthy aging. Recently, Yoo et al. [14] and Marques et al. [15] have proposed the analysis of the topology of the associated functional networks as an alternative way of quantifying CR. Interestingly, functional brain networks obtained during resting state show a positive correlation between CR and network efficiency [16,17]. Network efficiency, first introduced by Latora et al. [18] is a metric evaluating the inverse of the topological distance between nodes of a network. In the context of functional brain networks, the topological distance between two brain regions is normally defined as the inverse of their level of coordination. In this way, nodes that are not functionally connected (i.e., with zero coordination) have infinite distance between them. However, information between not connected regions (at an infinite topological distance) can travel indirectly through intermediate nodes. Network efficiency measures how close or distant the nodes of a network are in terms of their topological distance using the functional connections through intermediate nodes [18]. Under this framework, the positive correlation between CR and network efficiency during resting state relies on fMRI recordings, which report higher levels of synchronization between brain regions in individuals with higher CR[15,16]. However, during a cognitive process, the correlation between CR and synchronization turns to be negative, as shown in [11]. With the aim of clarifying whether CR is related to the topology of the functional networks during a memory task, we are concerned about the biomagnetic network profiles of two groups of healthy elders with different CR levels. Specifically, we obtained the MEG recordings of both groups during the performance of a modified Stenberg's test [17] and we explored the topological properties of the corresponding functional networks. Among the diversity of network metrics that one may extract from the functional networks we focused on the most extended ones: the network *strength*, which is a measure of the average level of synchronization along the functional network [19]; the *weighted clustering coefficient*, measuring the existence of triangles in the network and related to the local resilience [20]; the *eigenvector centrality* a measure of node importance obtained from the eigenvector associated to the largest eigenvalue of the adjacency matrix [21]; the average shortest path *d(i)*, measuring the average number of steps to travel from one node to another through the shortest topological distance [22]; the strength of the nearest neighbors *snn(i)*, indicating what is the average strength (i.e., sum of the weights of a node) of the neighbors (i.e., nodes directly connected) of a given node [19]; and the outreach *o(i)*, a network metric obtained by multiplying the weight of the connections by their Euclidean distance [23] (see Supp. Info. for a detailed definition of all network metrics).

We also quantified the dynamical properties of the brain regions in terms of entropy and complexity to compare both high and low CR groups. To the best of our knowledge, the present MEG network's approach, where the study of network metrics is combined with the dynamical properties of the nodes, has never been used within this field of research. Combining both approaches we have been able to detect differences between low and high CR individuals at the network level, showing that the network outreach and, consequently, the energetic expenses, is higher in individuals with low CR. At the same time, we show how the topological properties of the nodes (i.e., brain regions) are related to the complexity and entropy of their corresponding time series.

MATERIALS AND METHODS

*Participants*

The sample consisted of 20 healthy elders recruited from the "Geriatric Unit of the University Hospital San Carlos" (Madrid, Spain). All of them were right-handed [24], native Spanish speakers, and between 65 and 85 years old. Regarding the neuropsychological assessment, all of them met the following criteria: 1) a score above 28 in MMSE [25]; 2) no memory impairment as evidenced by delayed recall from Logical Memory II subtest of the Wechsler Memory Scale Revised (WMS-III-R; [26]); and 3) normal daily living activities measured by the Spanish version of the Functional Assessment Scale [27]. Exclusion criteria included: 1) a history of psychiatric or neurological disease; 2) psychoactive drugs consumption; and 3) severe sensory or comprehension deficits.

The 20 participants did not differ in physical, cognitive and social activities during late-life, and were further subdivided into two subgroups according to their CR level. The CR index (CRI) was estimated for each individual by adding the following two categories: 1) formal educational level (from 1 –illiterate/ functional illiterate- to 5- superior studies-) and 2) occupational attainment (from 1 –housewife- to 5 -business management/research-). According to their CRI score, 12 participants were classified into the low CR group if their CRI score was between 1 and 5; or into the high CR group (8 subjects) if their CRI score was between 6 and 10. In addition, both groups did not differ in age and in their performance during the cognitive task (see table 1).

Ethics statement: Methods were carried out in "accordance" with the approved guidelines. The investigation was approved by the local Ethics Committee (Madrid) and all participants signed a written informed consent before the MEG recordings.

|  | High- CR (n=8) | Low- CR (n=12) | P values |
| --- | --- | --- | --- |
| Age | 67.3±7.4 | 69.7±6.6 | $p>0.05$ |
| Educational Level | 4.5±0.7 | 2.4±0.7 | $p<0.05$* |
| Occupation attainment | 4.5±0.5 | 1.6±0.5 | $p<0.05$* |
| CRI score | 9±0.6 | 3.7±1.1 | $p<0.05$* |
| Cognitive activity | 4.2±2.3 | 2.8±1.4 | $p>0.05$ |
| Physical activity | 2.4±1.5 | 2.7±1 | $p>0.05$ |
| Social activity | 1.6±1.2 | 1.7±1 | $p>0.05$ |
| Acc. Smaqe | 114±10.7 | 106±14.1 | $p>0.05$ |

**Table 1**. Mean values ± standard deviation for high and low CR groups. Significant p values are p<0.05 and marked with an asterisk (*). Both groups differ in their educational level, occupation attainment and CRI scores, but not in the other variables. Acc. Smaqe, accuracy in the modified Stenberg's memory task.

*MEG task*

A modified version of the Sternberg's letter probe task was employed [17] (see Figure 1). Each subject was asked to remember a single set of 5 letters (i.e. 'SMAQE'). After that, there were presented a series of single letters, one at a time (500ms in duration with a random ISI between 2 and 3 seconds). They were asked to make a match/non-match decision by pressing a button with their right hand if the presented letter was one of the five initially memorized. The experiment consisted of 250 letters, in which half were initially presented and the rest were distracters, meaning they were not in the initial 5 letter set. All of the participants completed a training session until demonstrating they had correctly memorized the initial letter set. The task was projected through a LCD video projector (SONY VPLX600E) located outside the shielded MEG room through a series of mirrors. The screen was suspended approximately 1 meter above the participant's face. The letters subtended 1, 8 and 3 degrees of horizontal and vertical visual angle respectively.

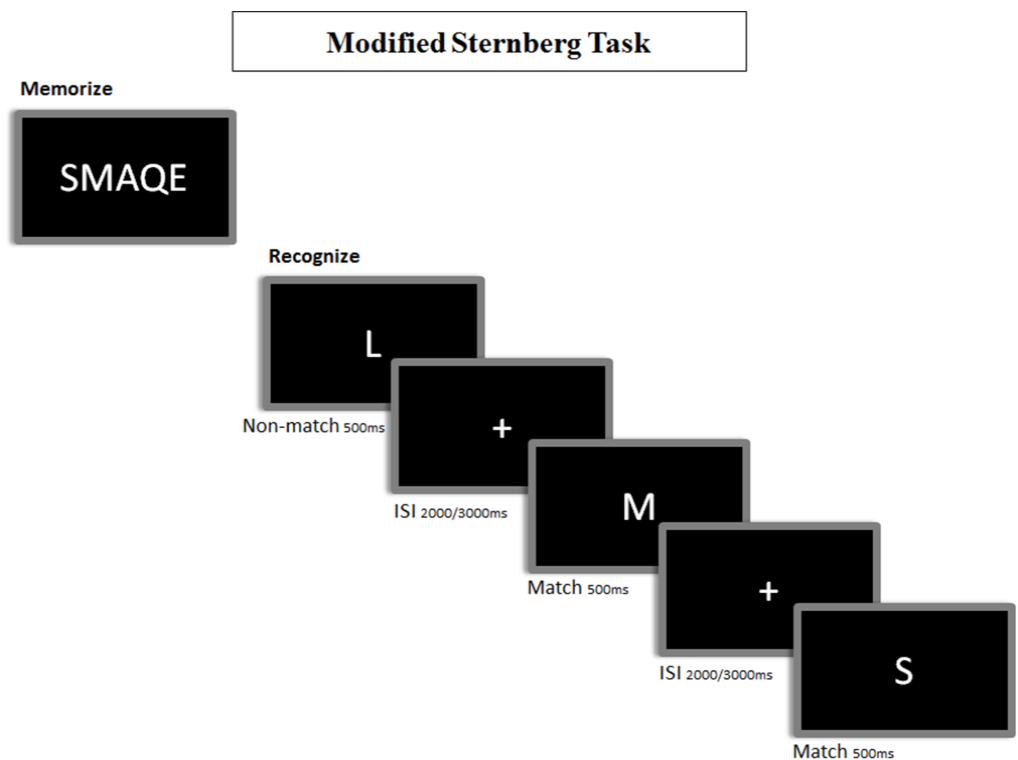

**Figure 1**. Representation of the modified version of the Sternberg's letter probe task. In the encoding phase, participants are instructed to memorize 5 letters (i.e.: "SMAQE"). In the recognition phase, participants are instructed to make a match/non-match button-press to indicate that the presented letter matched any of the encoded ones.

*MEG recordings*

MEG data was recorded during the execution of the modified Sternberg's task with a 254 Hz sampling rate and a band pass of 0.5 to 50 Hz, using a 148 channel whole-head magnetometer (MAGNES 2500 WH, 4D Neuroimaging) placed in a magnetically shielded room. In order to reduce external noise we employed an algorithm using reference channels at a distance from the MEG sensors. After trial segmentation, artifacts were visually inspected by an expert and epochs containing muscular artifacts or eye movements were discarded for further analysis. Only hits were considered, removing false alarms, correct rejections and omissions. We randomly selected a set of 35 trials for each individual, since it was the lowest number of successful epochs for all subjects in the study. Each trial consisted of each 230 time points for each of the 148 channels. Trials are not necessary consecutive, however, this issue is not crucial for the analysis of the coordination between sensors since each trial is analyzed individually and averaged later.

*Functional connectivity analysis*

To calculate the connectivity networks for each subject we used Synchronization Likelihood algorithm (SL) [28] which is a nonlinear measure of the synchronized activity that has been proven to be a suitable quantifier for datasets obtained from magnetoencephalographic recordings [29]. Specifically, the SL algorithm detects windows of repeated patterns within the time series of a channel A and, next, checks whether a channel B also shows a repeated pattern at the same time windows, no matter if it is the same or different to that observed in channel A. The range of values of SL is $0 \leq SL \leq 1$, being 0 when the time series of both channels are uncorrelated, and 1 for maximal synchronization [28]. SL was calculated for each pair of sensors since, in our study, sensors were the nodes of the functional networks. All pairs of *SL* are then included into a correlation matrix $W\{w_{ij}\}$ where $w_{ij}$ have values comprised between $\sim 0.05$ and $\sim 0.5$. We apply a linear normalization that leads to a probability matrix $P\{p_{ij}\}$, where the values of $p_{ij}$ are obtained as

$$p_{ij} = \frac{w_{ij} - \min[w_{ij}]}{\max[w_{ij}] - \min[w_{ij}]}. \qquad (1)$$

In this way, the probability matrix $P\{p_{ij}\}$, whose values are within the interval [0,1], reflects the probability of the presence of a link between nodes (brain regions) *i* and *j*.

*Graph metrics*

We calculated a series of metrics related with the role of the nodes within the network: the strength *s(i)*, the weighted clustering *cw(i)*, the eigenvector centrality *ec(i)*, the average shortest path *d(i)*, the strength of the nearest neighbours *snn(i)* and the outreach *o(i)*. We used the inverse of the values of the probability matrix $P\{p_{ij}\}$ for obtaining the "topological distances" between nodes to quantify the shortest paths. Another parameters such as the within-module degree *z(i)* (also kwon as the z-score) and the participation coefficient *p(i)* [23,30] were also computed using the community affiliation vector *Ccom(i)* extracted from the classical partition of the brain into six lobes: left-frontal, right-frontal, central, left-temporal, right-temporal and occipital. A series of global network features such as the global efficiency Eg and the average shortest path (d) were also calculated. The node average (average along the same node for all subjects within the same group) of several metrics was computed to obtain the following mean values: $\bar{s}(i)$, $\bar{c}_w(i)$, $\overline{ev}(i)$, $\bar{z}(i)$, $\bar{d}(i)$, ), $\bar{s}_{nn}(i)$ and $\bar{o}(i)$. Network averages of the preceding features were also appraised in order to compare both groups. Finally, we obtained the global efficiency $\bar{E}_g$ of the networks from the harmonic mean of the inverse of the shortest paths.

We constructed a set of randomized versions of the former functional networks, in order to evaluate to what extent the deviation of the network parameters between groups was a consequence of a topological reorganization or, on the contrary, just a matter of the number of links and their weights. With this purpose, we generated a group of 100 networks for each of the functional network. The randomization maintains the value of the links' weights by reshuffling the components of the weighted probability matrix $P\{p_{ij}\}$ (see [23] for a similar normalization). In this way, we guaranteed that the average strength of the network was maintained. Next, we calculated the network parameters for each of the randomized versions and obtained an average value of each metric. Finally, we normalized all network parameters with the average of the set of surrogate matrices, i.e., for a parameter **X** its normalized value would be:

$$\hat{X} = \frac{X}{X_{rand}} \qquad (2)$$

Note that this normalization allows focusing on changes related to the structure of the functional networks, since they exclude variations of the average weights of the networks (and their influence on the network metrics).

*Evaluation of the dynamical properties of the nodes*

In order to calculate the entropy and complexity brain signal, we used for each of the nodes in the network their corresponding time series comprising the 35 trials of the whole experiment. Since each trial has 230 time steps, we obtained times series of M = 8050 points for each node of each functional network.

Next, we applied the methodology introduced by [31] to quantify the dynamical properties of nonlinear dynamical systems. We calculated the normalized permutation entropy of the signal (S) as described in [32] to capture the level of uncertainty of a signal.

We calculated the complexity (H) of the signal for each subject following the procedure described in [33]. This measure is effective to quantify the complexity of different dynamical systems [34].

*Statistical analysis*

We performed permutation tests as well as non-parametric ones for all statistical hypotheses of this work. The permutation test is based on a t-statistical distance between two samples. For robustness, we constructed a distribution of 5000 different p-values associated to the number of permutations in each hypothesis. In each distribution, we found the percentile associated to the original p-value, thus dividing this percentile by the number of permutations as a correction of the resampling. We considered the rejection of the null hypothesis when the aforementioned ratio is bellow or equal to α = 0.05 level of significance. Results of permutation tests are presented throughout the paper. Complimentary, we performed a nonparametric Mann-Whitney U-test (rank-sum test) to show how the results do not change. The latter results are shown in the Supplementary Information.

RESULTS

*Micro-scale: Differences at the node level*

Firstly, we analysed those nodes with the highest eigenvector centrality in both groups ($\overline{ev}^{high,low}$). Figure 2 shows that both groups exhibited the most influencing nodes placed at the same cortical regions (i.e., at the occipital lobe). Nevertheless, we observed signs of a displacement of the node centrality toward the central and left-temporal lobes.

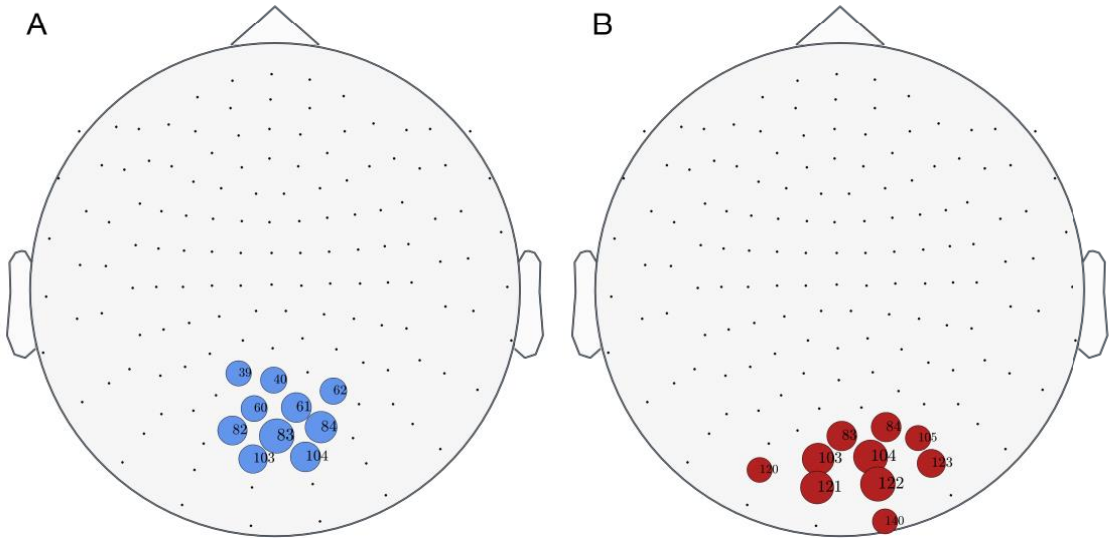

**Figure 2.** Back view for eigenvector centrality ($\overline{ev}(i)$) for high (A) and low (B) CR groups. Circles highlight the ten nodes with the highest $\overline{ev}(i)$. Note that they are localized close to the occipital lobe in the low CR group (blue circles) and slightly shift toward the central lobe for the high CR group (red circles).

In order to quantify the differences between groups, we obtained the average difference of the node centrality ($\Delta\overline{ev}(i)$), within-module degree $\Delta\bar{z}(i)$ and participation coefficient $\Delta\bar{p}(i)$ of each node for both groups.

Figure 3 shows the significant differences obtained in node centrality (A), within-module degree (B) and participation coefficient (C) between groups ($p_{val} \leq 0.05$). The high CR group showed higher values of the eigenvector centrality in nodes over the central lobe, near the parieto-occipital sulcus, as well as some tiny regions in the frontal lobe (i.e., $\overline{ev}(i)^{high} > \overline{ev}(i)^{low}$ and subsequently, $\Delta\overline{ev}(i) > 0$). Conversely, cortical tissue placed at the left-temporal and occipital lobes have higher centrality in the low CR population (i.e., $\Delta\overline{ev}(i) < 0$). High CR group presented higher within-module degree over both temporal brain areas and in the occipital lobe, while low CR group showed higher within-module degree in some regions located at the central lobe. Finally, there were few statistical differences between groups in the participation coefficient, showing the high CR group higher values than the low CR one, especially at regions placed at the frontal lobe. Statistical analysis based on a Man-Whitney U test shows similar results (see Supp. Info. for details).

It is worth mentioning that both the eigenvector centrality and the within-module degree are quantifying the percentage of importance a node has with regard to the other nodes of the network. Therefore, they should be interpreted as a way of evaluating the increase/decrease of importance with respect to all nodes, in the case of eigenvector centrality, or the nodes belonging to the same lobe, in the case of the within-module degree. Note that, if nodes belonging to the temporal and occipital lobes have higher centrality, it should be at expenses of other nodes of the functional network (in this case, those located at the central lobe), since the sum of the centralities of all nodes is constant (due to the fact that the modulus of the eigenvector is one). In other words, when comparing both groups, it would not be possible to observe higher centrality for all nodes of a given group (High CR or Low CR).

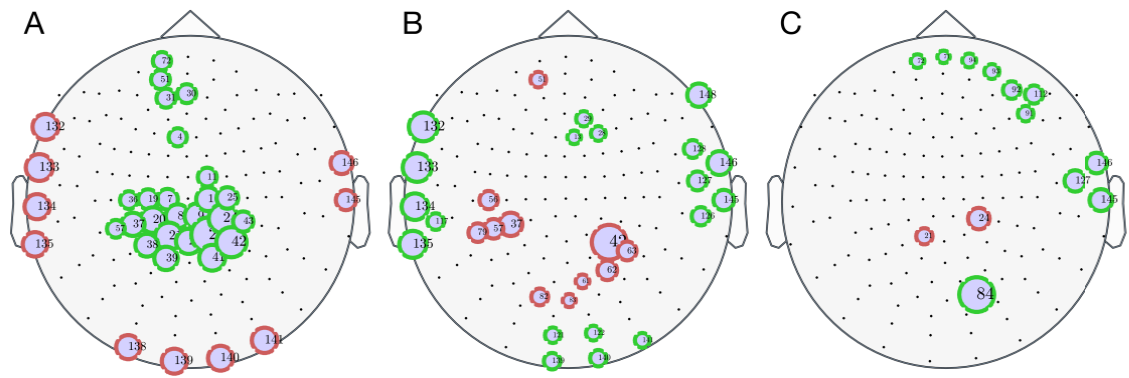

**Figure 3.** Differences between high and low CR groups at the node level. Black dots indicate the Euclidean position of the 148 magnetometers (nodes). Circles filled with lilac colour show nodes with significant statistical differences in: Eigenvector centrality $\Delta\overline{ev}(i)$ (**A**), within-module degree z-score $\Delta\overline{z}(i)$ (**B**) and participation coefficient $\Delta\overline{p}(i)$ (**C**). Green borders correspond to those nodes that have greater values in the high CR group, and red borders represent those nodes that exhibit greater values in the low CR group. Circle sizes are proportional to the absolute value of the differences between groups. Here we present the results obtained with the permutation test. Results obtained with Mann-Whitney U-test are similar (see Fig. S3 of Supp. Info. for details).

*Macro-scale: Analysing the topology of the whole network*

We analysed how the average network parameters were modified according to the level of CR of the participants. First, we calculated the mean and standard deviation of six network parameters for each subject (network strength $\bar{S}$, outreach $\bar{O}$, weighted clustering coefficient $\bar{C}_w$, average neighbour strength $\bar{S}_{nn}$, global efficiency $\bar{E}_g$ and average shortest path $\bar{d}$). Second, we carried out a permutation test and a nonparametric Mann-Whitney U-test (see Section Statistical analysis for details) for each network metric. Finally, those metrics with p-values<0.05 were accepted to have statistically significant differences.

In our results, the low CR group exhibited higher averages values than the high CR group in all parameters calculated, except for $\bar{d}$ (see Fig. 4 for details). However, only the network outreach $\bar{O}$ showed statistical significant differences both in the permutation (p= 0.0232) and the rank-sum (p=0.039) tests. Interestingly, the low CR group showed a higher $\bar{O}$ ($\bar{O}_{low}$= 13.57 and $\bar{O}_{high}$=12.29). It is also remarkable that the average shortest path length $\bar{d}$ exhibit significant differences (p=0.0491) at the rank sum test, despite not passing the permutation test. In this case the average values of shortest path in low CR are lower than in the high CR group ($\bar{d}_{low}$= 12.67 and $\bar{d}_{high}$=14.13) (see also Tab. S1 of the Supp. Info. for the values the averages).

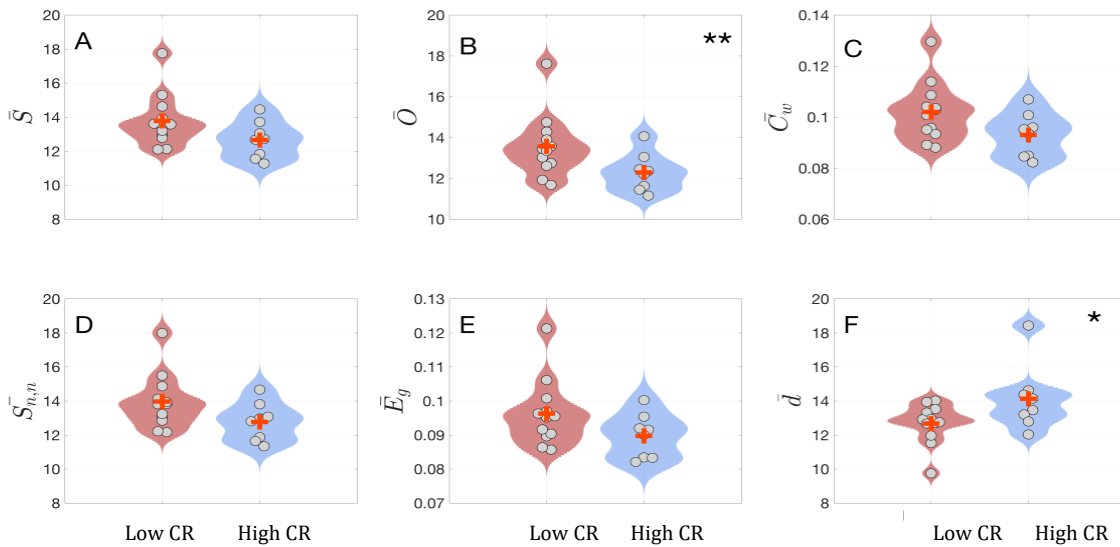

**Figure 4.** Violin plots showing the distribution of subject's features and means for the six network parameters analysed: Network strength $\bar{S}$, outreach $\bar{O}$, weighted clustering coefficient $\bar{C}_w$, average neighbour strength $\bar{S}_{nn}$, global efficiency $\bar{E}_g$ and average shortest path $\bar{d}$. Red and blue plots correspond to the low and high CR groups, respectively. Red crosses indicate the mean value. Outreach $\bar{O}$ (**B**) showed significant differences between groups both with the permutation test as well as with A Mann-Whitney U-test (indicated by two stars). Average shortest path $\bar{d}$ (**F**) showed statistical significant differences between groups only with the latter test (one star). Both statistical analyses were performed with $p_{val} < 0.05$.

*Dynamical Analysis of MEG Time Series*

We found a group of nodes showing statistical differences between the high and low CR groups, both for ***H(i)*** and ***C(i)*** ($p_{val} < 0.05$). In accordance with previous sections, the occipital lobe is the brain region that concentrates the highest amount of nodes with statistical differences (see Figures 5 and 6). We obtained a group of 23 nodes whose entropy was higher in the low CR group (except for one node) (see Figure 5). On the contrary, we obtained lower complexity of the signals in the low CR group (except for two nodes), but in a very similar brain localization that in the case of entropy (only two nodes did not overlap) (see Figure 6). These results indicate that a higher level of CR could be associated to a brain dynamics with lower entropy and higher complexity.

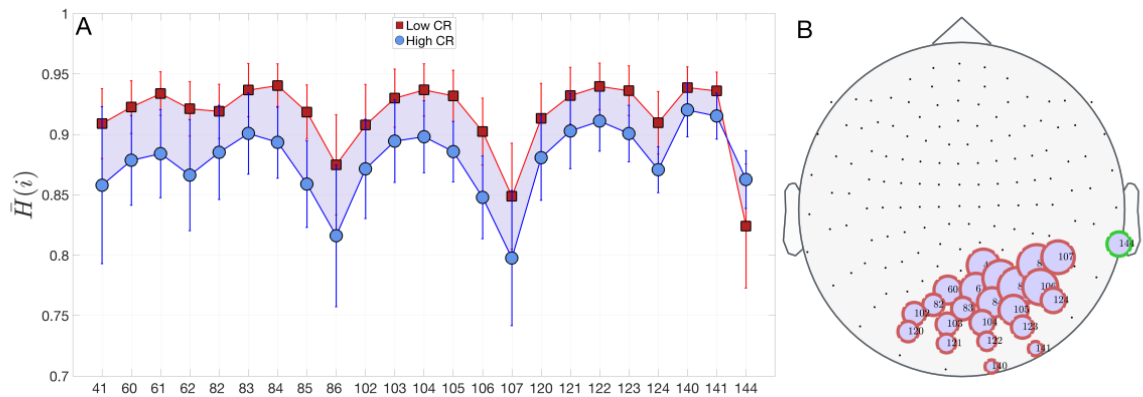

**Figure 5.** Differences of entropy $\bar{H}(i)$ between low and high CR groups. In A, we show the mean and standard deviations of the entropy of nodes that have statistical significant differences between groups. $\bar{H}(i)^{lowCR}$ (red squares) is higher than $\bar{H}(i)^{highCR}$ (blue circles) for nearly all nodes. In B, we plot the position of nodes with statistical differences. Node sizes are proportional to $\left| \Delta\bar{H}(i) = \bar{H}(i)^{highCR} - \bar{H}(i)^{lowCR} \right|$. Red borders represent $\Delta\bar{H}(i) < 0$, otherwise node borders are green. In this case, we used a permutation test but no relevant changes appear when a rank-sum test is carried out (see Supp. Info.).

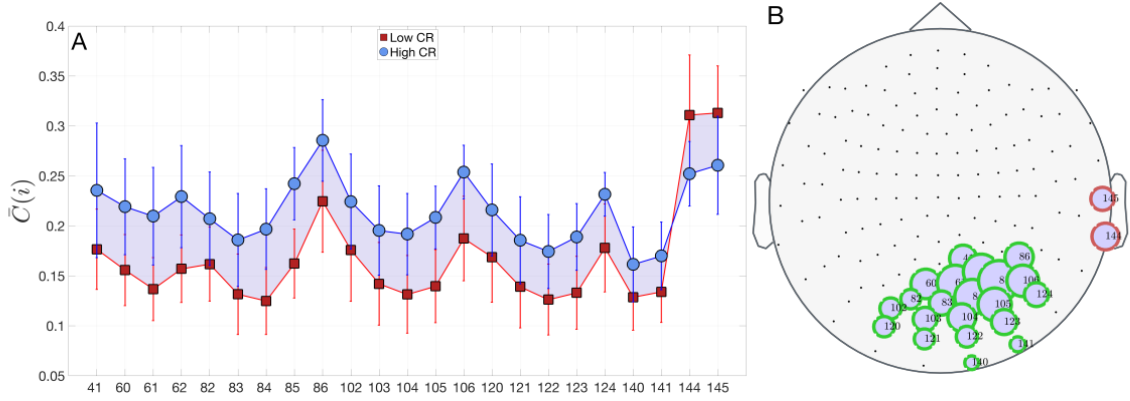

**Figure 6.** Differences of complexity $\bar{C}(i)$ between the Low and High CR groups. In A, we show the mean and standard deviations of the entropy of nodes that have statistical significant differences between groups. $\bar{C}(i)^{lowCR}$ (red squares) is lower than $\bar{C}(i)^{highCR}$ (blue circles) for nearly all nodes. In B, we plot the position of nodes with statistical differences. Node sizes are proportional to $\left| \Delta\bar{C}(i) = \bar{C}(i)^{highCR} - \bar{C}(i)^{lowCR} \right|$. Green borders represent $\Delta\bar{C}(i) > 0$, otherwise node borders are red. See Supp. Info. for similar results obtained with the rank-sum test.

*Correlation between Entropy and Complexity*

Then, we explored the relationship between these two dynamical properties. Figure 7 shows a complexity-entropy scatter plot for all nodes, with the inset plotting only those nodes with statistical differences. In the figure, each point corresponds to the average of each node over the subjects of a given group. We can observe a significant negative correlation between $H(i)$ and $C(i)$ for low ($r = 0.9874; p_{val} = 0.0010$) and high CR groups ($r = 0.8288; p_{val} r = 0.8288; p_{val} = 0.0010$), revealing that those nodes with less entropy are, at the same time, those nodes with higher complexity.

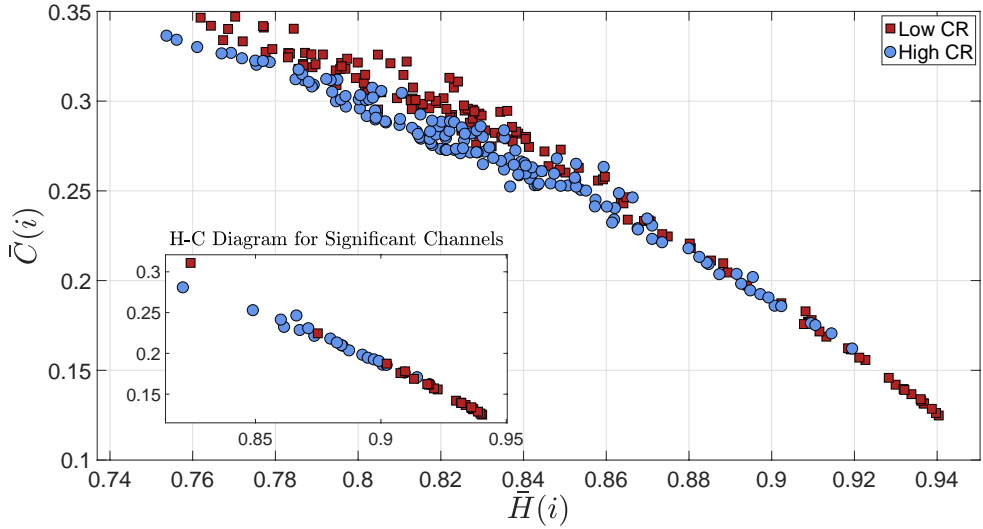

**Figure 7.** Complexity-Entropy Diagram. Diagram of subjects' average complexity vs. entropy for all 148 nodes (i.e., each point represents a node). Red squares correspond to low CR group and blue circles to the high CR group. The inset shows only nodes with statistical significant differences using the permutation test. Specifically, these nodes are the channels that have statistical differences when comparing both the mean entropies and complexities of high vs. low CR groups at node level. See Supp. Info. for similar results obtained with the rank-sum test.

*Correlations between Topology and Dynamics*

We explored the relationship between the dynamic properties of the nodes (i.e., $\bar{H}(i)$, $\bar{C}(i)$) and two of their topological parameters, namely the node strength $\bar{S}(i)$ and the weighted clustering coefficient $\bar{C}_w(i)$ for each CR group. Figure 8 shows all possible combinations.

For the low CR group, we found a positive correlation between $\bar{H}(i)$ and $\bar{S}(i)$ ($r^2 = 0.5833$; $p_{val} < 0.001$), which indicates that nodes with higher strength, i.e. the hubs of the network, are in turn those nodes with higher entropy at their dynamics (see Figure 8A). On the contrary, the correlation between the node complexity and its strength was negative ($r^2 = 0.7124$; $p_{val} < 0.001$) (see Figure 8B). The same pattern was found between $\bar{H}(i)$ and $\bar{C}_w(i)$ (positive correlation) ($r^2 = 0.5713$; $p_{val} < 0.001$), and between $\bar{C}(i)$ and $\bar{C}_w(i)$ (negative correlation) ($r^2 = 0.7055$; $p_{val} < 0.001$) (see Figure 8C-D).

Correlations found for the high CR group are qualitatively similar to the low CR but the values of the correlation parameter are systematically lower (See Tab. S2 of Supp. Info. for details).

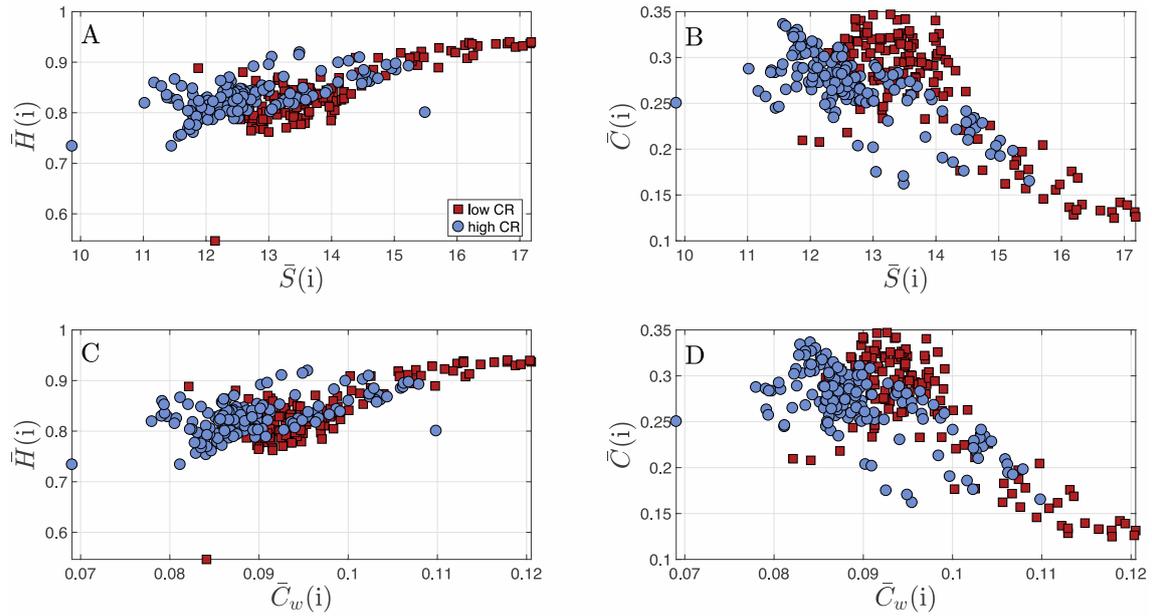

**Figure 8.** Correlation between the topological and dynamical metrics of the nodes. Correlation plots for the global dynamical properties (entropy $\overline{H}(i)$ and complexity $\overline{C}(i)$) of the MEG time series and their corresponding topological features (strength $\overline{S}(i)$ and clustering $\overline{C}(i)$). Specifically: A) $\overline{H}(i)$ vs. $\overline{S}(i)$; B) $\overline{C}(i)$ vs. $\overline{S}(i)$; C) $\overline{H}(i)$ vs. $\overline{C}_w(i)$ and D) $\overline{C}(i)$ vs. $\overline{C}_w(i)$. Red squares correspond to low CR group and blue circles to the high CR group. All statistical tests rejected the null hypothesis when using permutation test as well as with rank-sum test.

DISCUSSION

In the present study, we explored if CR might play a role in both the topology and the dynamics of the functional brain networks in healthy elders while performing a memory task. To this end, we classified the participants according to a CR index that combined educational level and occupational attainment into two groups: high or low CR level.

We firstly focused on the topological characteristics of the two groups, finding differences at microscale (i.e. nodes) and macroscale (i.e. average over all nodes) levels. At the microscale level (i.e., when nodes are analyzed), we observed that the most influential nodes for both groups were mainly placed over the occipital region, although there was a slight displacement towards the central lobe in the group with high CR. Specifically, we found that the low CR group exhibited higher eigenvector centrality over both temporal and occipital areas, while the high CR group presented higher values in central areas. These observations indicate that the importance of regions placed at the occipital and temporal regions is higher at the low CR group. At the same time, since the sum of the eigenvector centrality of all nodes of the network is constant, other nodes unavoidably have lower centrality values to compensate the higher values reported at the occipital and temporal regions. These nodes are placed at the central lobe, where the high CR group has higher centrality when compared with the low CR one. Interestingly, the within-module degree behaves in the opposite way. Comparing Fig. 3 A and B, we can observe that, in general, regions that have higher eigenvector centrality in the high CR group reduce their within-module degree. The explanation of this, in principle, counter-intuitive observation is that nodes have higher eigenvector centrality as a consequence of having higher weights at their long-range connections (i.e., in weight of the functional connections with nodes laying outside of their module).

Besides, high CR group showed higher participation coefficient values than the low CR group, especially in regions located at the frontal lobe. These results pointed out that the importance of a node and its connections with their neighbours inside and outside their community (lobe, in our case) may be differently located during the execution of a cognitive task, according to the level of CR.

At macroscale level, the low CR group exhibited higher values in different networks parameters compared to the high CR group (network strength, weighted clustering coefficient, average neighbour strength, global efficiency), being this increment statistically significant only in the outreach parameter both for the permutation and the rank-sum tests. Although strength results did not reach statistical significance, the low CR group presented a higher average in this network parameter suggesting that these participants may require higher brain synchronization to perform the memory task as well as the participants with high CR. As a result of this rise of functional connections, the average shortest path showed lower values, however, only the rank-sum test leaded to statistically significant differences between low and high CR groups. Along with the average shortest path, the outreach was the parameter more altered in the low CR group (in this case passing the two statistical tests). The higher outreach observed in the low CR group indicated the existence of a higher amount of long-range connections, implying a greater energetic cost to accomplish the task. Interestingly, this pattern was also described in a study carried out by our group [23], in which MCI and healthy subjects were compared performing the same Sternberg's letter-probe task. Network's results showed that MCI subjects exhibited an increment of strength and outreach values, and a decrement in the average shortest path, as formerly mentioned. Taken together all these findings, and especially the fact that the network outreach (and as a consequence, the energy consumption) is higher in the low CR group we may consider that brain functioning and network organization of those subjects with low CR have some similarities with MCI subjects during the execution of a memory task, and therefore we suggest that CR should be investigated as a possible factor with a remarkable impact on the cognitive status within the healthy aging. However, it is worth mentioning that, in our case, none of the normalized parameters showed significant statistical differences between low and high CR groups, while in MCI subjects showed differences in the normalized parameters indicating that their topology was more random than networks of the control group [23]. In our case, since the normalization is made by equivalent random networks, the lack of statistical differences between groups indicates that we can not claim that functional networks of the low CR group have a higher/lower random organization.

It is worth mentioning how the synchronized activity along the whole functional networks was negatively correlated with CR during the cognitive task [11]. In this sense, individuals with higher CR required less synchronization between brain regions to successfully carry out the memory test. On the contrary, fMRI recordings obtained during resting state reveal negative correlations, i.e., individuals with higher CR maintain higher synchronization in resting state [15,16]. This discrepancy suggests different mechanisms behind the creation and organization of the functional networks during resting state and cognitive tasks. Future studies should investigate this phenomenon with resting and task data with MEG.

The interpretation of the brain network profile described in the low CR group may be explained as the result of a neural compensatory effect [4]. In this case, as both groups did not differ in their memory task accuracy, we may consider it as a successful neural compensation since the low CR group had to engage additional resources to maintain or improve their performance [12]. This effect may be related to the Compensation-Related Utilization of Neural Circuits Hypothesis (CRUNCH) [35], which proposes that there is a loss of efficiency in brain networks of aging, when compared with young adults, since additional brain areas are needed to perform cognitive tasks. In the present study, we used a low demanding memory task, but if we would had modified its difficulty, we would had probably found an unsuccessful compensation attempt in the low CR group, as previously described in MCI subjects [36]. The cause of this compensatory effect may be the beginning of a network malfunctioning since individuals with low CR exhibited a lower energetic efficiency to perform the memory task, as previously described in MCI subjects [23]. In this line, these individuals may exhibit clinical symptoms of pathological aging (i.e. MCI) earlier or more severe than those subjects with a higher CR.

Another noticeable novelty explored in the present work was the dynamical properties of the cortical brain regions while performing the memory task. In line with the microscale results, the main differences between both groups were located over the parieto-occipital lobe. In fact, a lower CR level was related with lower complexity and higher entropy. This means that the brain signals of the low CR group presented a pattern more disorganized than the high CR group within the occipital lobe. In addition, the low CR group exhibited the opposite pattern in the right temporal lobe (i.e. less entropy and more complexity). We then may hypothesize that the recruitment of posterior brain areas was not providing an efficient contribution to adequately perform the task, and therefore the individuals with low CR needed to involve right temporal regions to achieved it. We then verified that these two dynamical properties were inversely correlated, independently of the CR level. Finally, one of the most original approaches performed in this study was the exploration of the relationship between the topology (i.e. node strength and weighted clustering coefficient) and the dynamic characteristics of the nodes (i.e. entropy and complexity). We found that in both CR groups, greater values in both topological measures were related with a higher entropy and a lower complexity. These results demonstrated that the most important nodes or hubs within the functional brain networks were those that exhibited more random dynamics, no matter what the level of CR of the individuals.

The results obtained in the current study are limited due to the lack of follow-up of the participants and the small sample size, especially in the case of the high CR group. Notwithstanding, it still demonstrated that CR plays a relevant role in the topological and dynamical network's properties in healthy aging. It should be pointed that one of the advantages of conducting the analysis in the sensor space was that MEG signal was not manipulated, affecting neither the complexity nor the entropy. In addition, these results corroborated that network analysis may be a suitable approach for studying CR, and that MEG may be an appropriate tool to detect functional brain changes in healthy aging. Thus, the profiles described in this work for the low CR group indicate a dual pattern of compensation and network impairment. As a consequence, if one of these subjects suffers brain damage or starts a neurodegenerative disease process, their closeness to random network organization would probably be a risk factor for developing a more severe cognitive impairment, although some compensation mechanisms are still present. Thus, this study highlights the importance of preventive interventions, including cognitive training, which could induce a more protective brain network organization in the case of brain damage in elderly subjects.

## ACKNOWLEDGMENTS

This study was supported by the Spanish Ministry of Economy and Competitiveness under Project FIS2013-41057-P, a postdoctoral fellowship from the Complutense University of Madrid to María Eugenia López and a predoctoral fellowship to David López Sanz (BES-2013-063772), both from the Spanish Ministry of Economy and Competitiveness.

## AUTHOR CONTRIBUTIONS

JHM contributed writing paper, in-silico experimental setup, wrote code for results, analysed the data; MEL with MEG recordings and writing paper; PA analysed the data; MC contributed with results and statistical analysis; JAP-P worked on the statistical analysis; DL-S contributed writing the paper; PG worked on the clinical assessment of the participants, FM helped in the design the paradigm and JMB conceived the analysis, analyzed the data and helped writing the manuscript.

## ADDITIONAL INFORMATION

Competing financial interests: The authors declare no competing financial interests.

## REFERENCES

1.  Stern, Y. Cognitive reserve in ageing and Alzheimer's disease. *Lancet Neurol.* **11,** 1006–12 (2012).
2.  Katzman, R. Education and the prevalence of dementia and Alzheimer's disease. *Neurology* **43,** 13–20 (1993).
3.  Satz, P. Brain reserve capacity on symptom onset after brain injury: A formulation and review of evidence for threshold theory. *Neuropsychology* (1993). doi:10.1037/0894-4105.7.3.273
4.  Stern, Y. Cognitive reserve. *Neuropsychologia* **47,** 2015–2028 (2009).
5.  Valenzuela, M. J. & Sachdev, P. Brain reserve and dementia: A systematic review. *Psychol. Med.* **36,** 441–454 (2006).
6.  Bartrés-Faz, D. & Arenaza-Urquijo, E. M. Structural and functional imaging correlates of cognitive and brain reserve hypotheses in healthy and pathological aging. *Brain Topogr.* **24,** 340–57 (2011).
7.  Serra, L. *et al.* Neuroanatomical correlates of cognitive reserve in Alzheimer disease. *Rejuvenation Res.* **14,** 143–51 (2011).
8.  Perneczky, R. *et al.* Schooling mediates brain reserve in Alzheimer's disease: findings of fluoro-deoxy-glucose-positron emission tomography. *J. Neurol. Neurosurg. Psychiatry* **77,** 1060–1063 (2006).
9.  Serra, L. *et al.* Cognitive reserve and the risk for Alzheimer's disease: a longitudinal study. *Neurobiol. Aging* **36,** 592–600 (2015).
10. López, M. E. *et al.* Searching for Primary Predictors of Conversion from Mild Cognitive Impairment to Alzheimer's Disease: A Multivariate Follow-Up Study. *J. Alzheimers. Dis.*

**52,** 133–143 (2016).

11. Solé-Padullés, C. *et al.* Brain structure and function related to cognitive reserve variables in normal aging, mild cognitive impairment and Alzheimer's disease. *Neurobiol. Aging* **30,** 1114–24 (2009).

12. López, M. E. *et al.* Cognitive reserve is associated with the functional organization of the brain in healthy aging: A MEG study. *Front. Aging Neurosci.* **6,** 1–9 (2014).

13. Bullmore, E. & Sporns, O. Complex brain networks: graph theoretical analysis of structural and functional systems. *Nat. Rev. Neurosci.* **10,** 186–98 (2009).

14. Wook Yoo, S. *et al.* A Network Flow-based Analysis of Cognitive Reserve in Normal Ageing and Alzheimer's Disease. *Sci. Rep.* **5,** 10057 (2015).

15. Marques, P. *et al.* The functional connectome of cognitive reserve. *Hum. Brain Mapp.* **37,** 3310–3322 (2016).

16. van den Heuvel, M. P., Mandl, R. C. W., Kahn, R. S. & Hulshoff Pol, H. E. Functionally linked resting-state networks reflect the underlying structural connectivity architecture of the human brain. *Hum. Brain Mapp.* **30,** 3127–41 (2009).

17. deToledo-Morrell, L. *et al.* A 'stress' test for memory dysfunction. Electrophysiologic manifestations of early Alzheimer's disease. *Arch. Neurol.* **48,** 605–9 (1991).

18. Latora, V. & Marchiori, M. Efficient behavior of small-world networks. *Phys. Rev. Lett.* **87,** 198701 (2001).

19. Newman, M. E. J. (Mark E. J. . *Networks : an introduction*. (Oxford University Press, 2010).

20. Ahnert, S. E., Garlaschelli, D., Fink, T. M. A. & Caldarelli, G. Ensemble approach to the analysis of weighted networks. *Phys. Rev. E* **76,** 16101 (2007).

21. Navas, A. *et al.* Functional Hubs in Mild Cognitive Impairment. *Int. J. Bifurc. Chaos* **25,** 1550034 (2015).

22. Watts, D. J. & Strogatz, S. H. Collective dynamics of 'small-world' networks. *Nature* **393,** 440–2 (1998).

23. Buldú, J. M. *et al.* Reorganization of functional networks in mild cognitive impairment. *PLoS One* **6,** e19584 (2011).

24. Oldfield, R. C. The assessment and analysis of handedness: the Edinburgh inventory. *Neuropsychologia* **9,** 97–113 (1971).

25. Lobo, A., Ezquerra, J., Gómez Burgada, F., Sala, J. M. & Seva Díaz, A. [Cognocitive mini-test (a simple practical test to detect intellectual changes in medical patients)]. *Actas Luso. Esp. Neurol. Psiquiatr. Cienc. Afines* **7,** 189–202 (1979).

26. Wechsler, D. *Wechsler Memory Scale - Revised*. (The Psychological Corporation, 1987).

27. Pfeffer, R. I., Kurosaki, T. T., Harrah, C. H., Chance, J. M. & Filos, S. Measurement of functional activities in older adults in the community. *J. Gerontol.* **37,** 323–9 (1982).

28. Stam, C. J. & van Dijk, B. W. Synchronization likelihood: an unbiased measure of generalized synchronization in multivariate data sets. *Phys. D Nonlinear Phenom.* **163,** 236–251 (2002).

29. Stam, C. J. *et al.* Magnetoencephalographic evaluation of resting-state functional connectivity in Alzheimer's disease. *Neuroimage* **32,** 1335–44 (2006).

30. Guimerà, R. & Amaral, L. A. N. Cartography of complex networks: modules and universal roles. *J. Stat. Mech. Theory Exp.* **2005,** P02001 (2005).

31. Bandt, C. & Pompe, B. Permutation entropy: a natural complexity measure for time series. *Phys. Rev. Lett.* **88,** 174102 (2002).

32. Rosso, O. A., Larrondo, H. A., Martin, M. T., Plastino, A. & Fuentes, M. A. Distinguishing noise from chaos. *Phys. Rev. Lett.* **99,** 154102 (2007).

33. López-Ruiz, R., Mancini, H. L. & Calbet, X. A statistical measure of complexity. *Phys. Lett. A* **209,** 321–326 (1995).

34. Zanin, M., Zunino, L., Rosso, O. A. & Papo, D. Permutation Entropy and Its Main Biomedical and Econophysics Applications: A Review. *Entropy* **14,** 1553–1577 (2012).

35. Reuter-Lorenz, P. a. & Cappell, K. a. Neurocognitive aging and the compensation hypothesis. *Curr. Dir. Psychol. Sci.* **17,** 177–182 (2008).

36. López, M. E. *et al.* Synchronization during an internally directed cognitive state\nin healthy aging and mild cognitive impairment: a MEG study. *Age (Omaha).* **36,** 9643 (2014).

# SUPLEMENTARY INFORMATION

## S1.- DEFINITION OF NETWORK METRICS

The strength *s(i)* of a node *i* is the sum of the weights $w_{ij}$ of all its links:

$$s(i) = \sum_{j \in N} w_{ij}$$

The strength *S* of a functional network is the average of the strength of all its nodes.

The strength of the nearest neighbours *snn(i)* as the average strength of all neighbours of node *i*.

The outreach *o(i)* of a node *i* is the sum of the links' weights $w_{ij}$ multiplied by the link Euclidean lengths $l_{ij}$ between node *i* and node *j*:

$$o(i) = \sum_{j \in N} l_{ij} w_{ij}$$

The outreach *O* of a functional network is the average of the outreach of all its nodes.

The weighted clustering coefficient $c_w(i)$ of a node *i* quantifies the percentage of neighbours of a certain node that, in turn, are neighbours between them, taking into account the weight of the connections:

$$c_w(i) = \frac{\sum_{jk} w_{ij} w_{jk} w_{ik}}{\sum_{jk} w_{ij} w_{ik}}$$

The weighted clustering coefficient $C_w$ of a functional network is the average of the weighted clustering coefficient of all its nodes.

The eigenvector centrality *ev(i)* of a node *i* measures the importance of a node according to the importance of its neighbours. It is obtained from the eigenvector associated with the largest eigenvalue of the connectivity matrix.

The within-module degree *z(i)* of a node *i* measures the importance of a node inside its community. In our case, the community of a node is the lobe the node belongs to:

$$z(i) = \frac{k_i(m_i) - \langle k_i(m_i) \rangle}{\sigma_{k(m_i)}}$$

where $k_i(m_i)$ is the degree of node *i* inside its community (lobe), and $\langle k_i(m_i) \rangle$ and $\sigma_{k(m_i)}$ are the average and the standard deviation of the degree inside the community, respectively.

The participation coefficient *p(i)* of node *i* quantifies the percentage of links of a node that reach other communities:

$$p(i) = 1 - \sum_m \left(\frac{k_i(m)}{k_i}\right)^2$$

where $k_i(m)$ is the degree of node *i* inside community *m*.

The average shortest path *d* of node *i* is the average of the minimum number of nodes to be visited when going from node *i* to *j*. To obtain *d*, we weight the distances between nodes $D_{ij}$, as the inverse of the elements of the connectivity matrix $w_{ij}$, i.e. $D_{ij} = 1/w_{ij}$. Next, we calculate the shortest-path distance between every pair of nodes using the Dijkstra's algorithm (Dijkstra, 1959). This way, we obtain the shortest-path matrix $dis_{ij}$ (Newman, 2010), and finally the average shortest path *d* is obtained as the average of the distance of each node to the rest of the network:

$$d = \frac{1}{N(N-1)} \sum_{i \neq j} dis_{ij}$$

The global Efficiency *Eg* of a network, first introduced by Latora & Marchiori (2001), overcomes the fact that certain nodes of a network could be isolated from the others, thus leading to infinite distance between them. Mathematically, *Eg* is defined as the harmonic mean of the inverse of the shortest paths between all nodes of the network, with $dis_{ij}$, being the shortest path between nodes i and j:

$$E_g = \frac{1}{N(N-1)} \sum_{i \neq j} \frac{1}{dis_{ij}}$$

## S2.- ENTROPY AND COMPLEXITY

In the matter of dynamical complexity, information plays an important role as a feature that describes the outermost bounds of periodicity, chaos and complexity. In this sense, the Bandt & Pompe (2002) BP method obtains the intrinsic temporal symbol sequences *{St}* from the neighbouring steps of a time series (see Fig. S1 for a qualitative explanation). This symbol sequences depend on an embedding dimension *D* = 3, 4, 5, …, which represents the amount of past information, being *D* the number of neighbouring samples. In this way, *D* characterizes each

*{Xt}* time series along *t* = 1, 2, ..., *M* samples. To do that, *{Xt}* is partitioned into (*M* - *D*) overlapping vectors of dimension *D*.

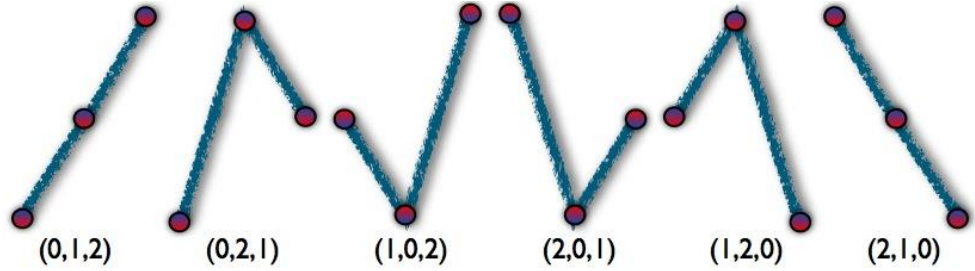

**Figure S1:** Overlapping vectors for the case *D* = 3. *D!* is the number of patterns. This way 3! different types π of accessible states are presented. The probability of appearance of the ordinal patterns is contained in *P*. If an ordinal pattern never appears, it is called a forbidden pattern.

The greater the dimension *D*, the more information about the past state of our system, and the longer the vectors are (containing the ordering of a set of *D* samples). Each of the vectors is assigned to a time *t*, sliding the vector at every time step, to get a total of (*M* – *D* + *1*) overlapping vectors. Hence, for each (*M* - *D*) vector, the position of the lowest value will be assigned the ordinal value zero. The position of the highest value will correspond to (*D* - *1*) ordinal value (the highest in the ranking). Thus, the following positions in-between the assigned zero and (*D* - *1*) will be assigned by rating the positions of the remaining samples in the respective ordinal values. When all (*M* – *D* + *1*) different order types in *{St}* are calculated, it is possible to obtain the probability distribution function (PDF) *P(π)*, quantifying the probability of finding a certain order pattern associated to *{Xt}*:

$$P(\pi) = \frac{\#\{t | t \leq M - D, (x_{t+1}, \ldots, x_{t+D}) \text{ has a type } \pi\}}{M - D + 1}$$

In previous equation, π is a possible ordinal pattern presented in the sequence *{St}* and # is its number of appearances. Note that each ordinal pattern is a permutation of π = (0,1,2, ..., *D* - 1).

In other words, *D!* represents all possible permutations π of order *D* of the number of accessible states (M - D). As an example, consider the case of *D* = 3 in Fig. 1.8. The number of patterns or accessible states will be *D!* = *3!* = *6* and the possible patterns π will be: {(012), (021), (102), (201), (120), (210)}. From the former ones, vectors that appear in *{St}* are called ordinal patterns of *{Xt}*, those that do not appear in *{St}*, but belong to the possible accessible states are called forbidden patterns. Finally, the discrete PDF of the ordinal *patterns P =$p_j$, ∀ j = 1,2, ..., N ∧ N = D!* is calculated. This PDF, obtained from BP method, carries the temporal information of *{Xt}* by comparing consecutive samples. In other words, this symbolic technique incorporates the causality effects of a short-memory (of *D* steps) a time series has.

Next, we use the PDF of the ordinal patterns to define the *Normalized Permutation Entropy H* (Band & Pompe, 2002):

*Definition S1 — Normalized Permutation Entropy H[P]. It is given by the ratio between the entropy S[P] of the ordinal patterns and Smax = S[Pe], being Pe the uniform probability distribution:*

$H[P] = S[P]/S_{max}$

Note that, the normalized permutation entropy *H[P]* is bounded between [0, 1].

Regarding the finite size effects, the normalized permutation entropy *H* allows to include a uniform distribution *Pe* = {1/N, 1/N, …, 1/N} making *H* to be an intensive property. This uniform distribution *Pe*, also maximizes the associate-system information entropy *S[P]*, i.e., *Smax =log(N)= log(D!)*. This way, the amount of disorder *H[P]* based on the information measure *S[P]* associated to *P* is defined as the Permutation Entropy because it runs over all *D!* permutations $\pi$ of order *D*.

On the other hand, the insertion of "a priori" equilibrium distribution {*Pe*} as a correction for the associated entropy, leads to a discrimination between two populations. In other words, we need to evaluate the distance between both distributions *P* and *Pe*. This fact makes *S[P]* not being enough to effectively characterize *{Xt}* because there could be some ordinal patterns that belong to *P* as well as to *Pe*. This distance accounts for the "order" of the system when one of the few ordinal patterns emerges as the preferred one. The disequilibrium between statistical populations will be the measure to distinguish this non-Euclidean distance. We quantify the disequilibrium *Q* by adopting some statistical distance *D* between the possible and accessible states of the systems in *P* and the equilibrium distribution *Pe*:

*Definition S2 — Disequilibrium Q[P]. It evaluates the distance between P and Pe as:*

$Q[P] = Q_0 \, D[P, Pe]$

where $Q_0$ is a normalization constant leading to $0 \leq Q[P] \leq 1$. A $Q \neq 0$ indicates the existence of preferred states among the accessible ones.

In this way, the disequilibrium *Q[P]*, discriminates ordinal patterns in *P* from the uniform distribution *Pe*. The zero limit or the minimum disequilibrium, implies that the lowest separation of both populations does not distinguish between ordinal patterns coming from both populations. Meanwhile the upper limit, with a high disequilibrium, is related to the fact of the existence of some privileged ordinal patterns in *P*.

Hitherto, both *H* and *Q* give some sense of the understanding of what the dynamical properties of the system are. Nevertheless, we are concerned on evaluating the interplay between the order and disorder of a system. Therefore, it is also desirable to complement these measures with some metric quantifying the complexity of the system.

In this way, Bandt and Pompe define the statistical complexity of a system as:

*Definition S3 — Statistical Complexity C[P]. This complexity mesures is defined as the product between the permutation entropy H and the disequilibrium Q:*

$C[P] = H[P] * Q[P]$

With this definition, the statistical complexity $C$ accomplishes the first requirement since $H$ and $Q$ are intensive quantities. The second requirement is also achieved since, by means of $H$ and $Q$, we are measuring the disorder of a system and its distance from the equilibrium. Note that the statistical complexity vanishes either if the system is at equilibrium (maximum disorder) or if it is completely ordered (maximal distance from the equilibrium). Figure 1.9 shows a qualitative plot indicating the interplay between the three measures.

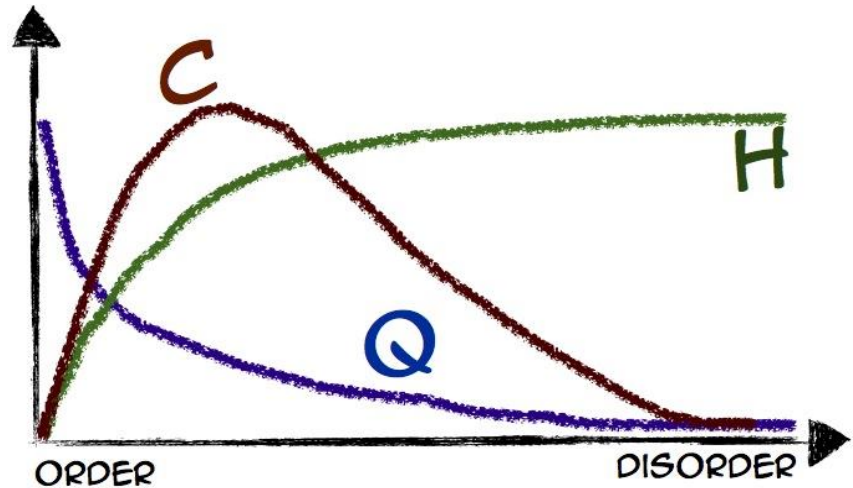

**Figure S2:** Statistical measures based on ordinal patterns. Schematic representation of the interplay between the normalized permutation entropy *H[P]*, disequilibrium *Q[P]* and the statistical complexity *C[P]*, in terms of a system that range from complete order to complete disorder.

The previous definitions of *H*, *Q* and *C* are usually known as Generalized Statistical Complexity Measures (SCM). SCM capture either, the essential details of the dynamics that allow discerning among different degrees of periodicity and randomness, as well as all possible degrees of stochasticity when the information of *{Xt}* is extracted via the BP method. SCM, not only compute randomness, but a wide range of correlation structures, not already offered by a simple entropy analysis.

**S3.- COMPLEMENTARY RESULTS**

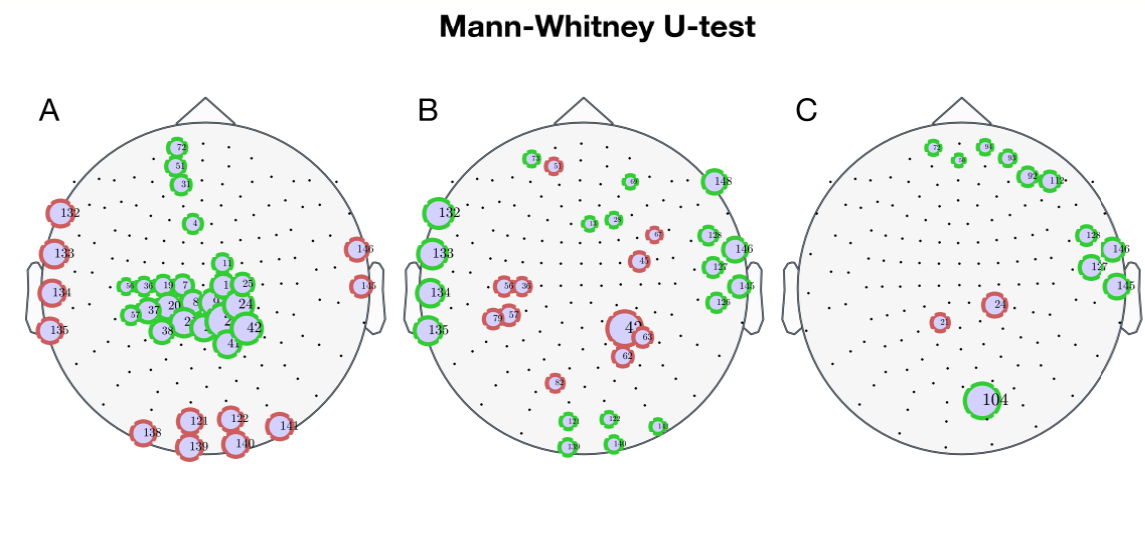

**Figure S3.** Differences between high and low CR groups at the node level. Black dots indicate the Euclidean position of the 148 magnetometers (nodes). Circles filled with lilac colour show nodes with significant statistical differences obtained with Mann-Whitney U-test in the: Eigenvector centrality $\Delta\overline{ev}(i)$ (**A**), within-module degree z-score $\Delta\bar{z}(i)$ (**B**) and participation coefficient $\Delta\bar{p}(i)$ (**C**). Green borders correspond to those nodes that have greater values in the high CR group, and red borders represent those nodes that exhibit greater values in the low CR group. Circle sizes are proportional to the absolute value of the differences between groups. Here we present the results obtained with the permutation test.

| Group | $\bar{S}$ | $\bar{O}$ (**) | $\bar{C}_w$ | $\bar{S}_{nn}$ | $\bar{E}_g$ | $\bar{d}$ (*) |
|---|---|---|---|---|---|---|
| *Low* | 13.78 | 13.57 | 0.102 | 13.96 | 0.096 | 12.69 |
| *High* | 12.75 | 12.29 | 0.093 | 12.76 | 0.089 | 14.12 |

**Table S1.** Average network parameters of the low and high CR groups. Specifically, the network strength $\bar{S}$, the outreach $\bar{O}$, the weighted clustering coefficient $\bar{C}_w$, the average neighbour strength $\bar{S}_{nn}$, the global efficiency $\bar{E}_g$ and the average shortest path $\bar{d}$. Asterisks indicate those metrics with statistically significant differences between groups: one asterisk for the parameters passing the rank-sum test and two asterisks for those parameters that also passed the permutation test.

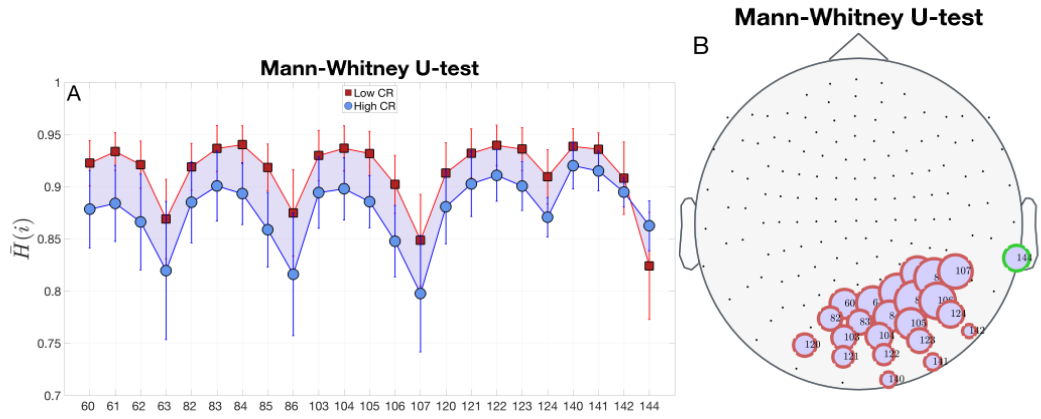

**Figure S4.** Differences of entropy $\bar{H}(i)$ between low and high CR groups. In A, we show the mean and standard deviations of the entropy of nodes that have statistical significant differences between groups. $\bar{H}(i)^{lowCR}$ (red squares) is higher than $\bar{H}(i)^{highCR}$ (blue circles) for nearly all nodes. In B, we plot the position of nodes with statistical differences. Node sizes are proportional to $|\Delta\bar{H}(i) = \bar{H}(i)^{highCR} - \bar{H}(i)^{lowCR}|$. Red borders represent $\Delta\bar{H}(i) < 0$, otherwise node borders are green. In this case, significant statistical differences were obtained with Mann-Whitney U-test.

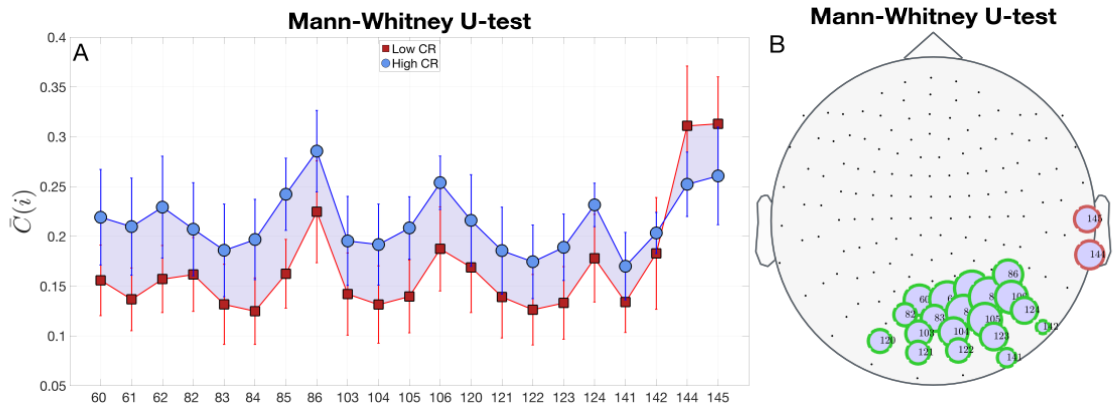

**Figure S5.** Differences of complexity $\bar{C}(i)$ between the Low and High CR groups. In A, we show the mean and standard deviations of the entropy of nodes that have statistical significant differences between groups. $\bar{C}(i)^{lowCR}$ (red squares) is lower than $\bar{C}(i)^{highCR}$ (blue circles) for nearly all nodes. In B, we plot the position of nodes with statistical differences. Node sizes are proportional to $|\Delta\bar{C}(i) = \bar{C}(i)^{highCR} - \bar{C}(i)^{lowCR}|$. Green borders represent $\Delta\bar{C}(i) > 0$, otherwise node borders are red. Statistical significant differences were obtained with Mann-Whitney U-test.

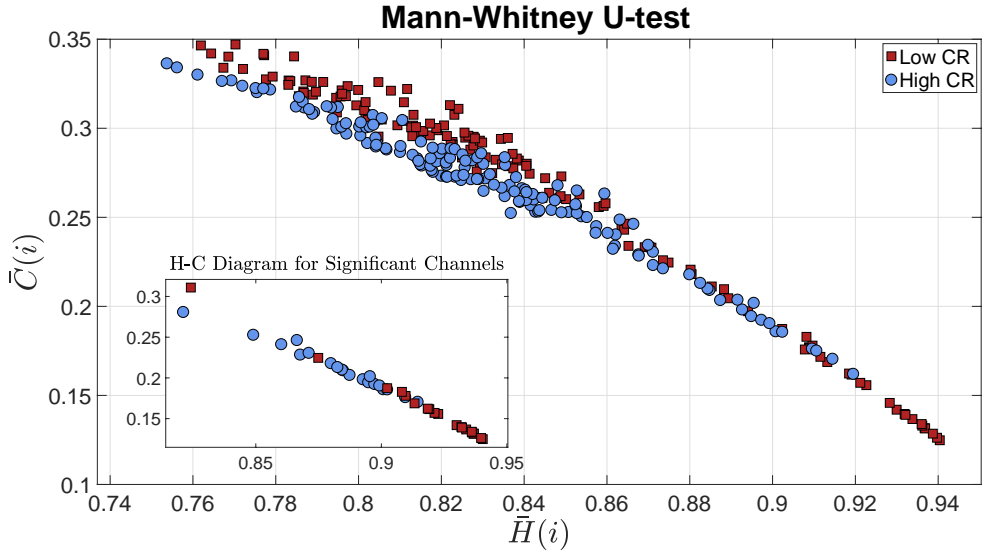

**Figure S6.** Complexity-Entropy Diagram. Diagram of subjects' average complexity vs. entropy for all 148 nodes (i.e., each point represents a node). Red squares correspond to low CR group and blue circles to the high CR group. The inset shows only nodes with statistical significant differences using the permutation test. Specifically, these nodes are the channels that have statistical differences when comparing both the mean entropies and complexities of high vs. low CR groups at node level. Statistical significant differences were obtained with Mann-Whitney U-test.

| $r^2$ | $(\bar{S}, \bar{H})^{Low}$ | $(\bar{S}, \bar{H})^{High}$ | $(\bar{S}, \bar{C})^{Low}$ | $(\bar{S}, \bar{C})^{High}$ | $(\bar{C}_w, \bar{H})^{Low}$ | $(\bar{C}_w, \bar{H})^{High}$ | $(\bar{C}_w, \bar{C})^{Low}$ | $(\bar{C}_w, \bar{C})^{High}$ |
|---|---|---|---|---|---|---|---|---|
| Lin. | 0.5833 | 0.4105 | 0.6486 | 0.4902 | 0.5719 | 0.3423 | 0.6417 | 0.4246 |
| Pol. | 0.5833 | 0.4075 | 0.7124 | 0.5159 | 0.5713 | 0.3332 | 0.7055 | 0.4733 |

**Table S2.** Coefficient $r^2$ for the correlations of Fig. 8 of the main text. First (second) row represent the linear (second order polynomic) fit. Columns represent the $r^2$ of both averaged variables (structural and dynamical) for the low and high CR groups. Columns 1 and 2 are for Fig. 8A, columns 3 and 4 are for Fig. 8B, columns 5 and 6 are for Fig. 8C. The last two columns are for Fig. 8D. Permutation test and rank-sum test rejected Ho with p<0.001.

**REFERENCES**

Bandt, C. & Pompe, B. Permutation Entropy: A Natural Complexity Measure for Time Series. *Phys. Rev. Lett.* **88**, 174102 (2002).

Dijkstra, E.W. A note on two problems in connexion with graphs. *Numerische Mathematik* **1**, 269–271 (1959).

Latora, V. & Marchiori, M. Efficient behavior of small-world networks. *Physical Review Letters* **87**, 198701 (2001).

Newman, M. E. J. Networks: An introduction. (Oxford University Press, 2010).

Garibotto,V., Borroni,B., Sorbi,S., Cappa,S.F., Padovani,A., and Perani,D.(2011). Education and occupation provide reserve in both ApoE ε4 carrier and non carrier patients with probable Alzheimer's disease. Neurol.Sci. 33, 1037–1042. doi:10.1007/s10072-011-0889-5